\documentstyle[12pt]{article}
\begin{document}
\begin {center}
{\bf
{\Large $ \omega (\to \pi^+\pi^-\pi^0) $ meson photoproduction on proton }
}
\end {center}
\begin {center}
Swapan Das  \\
{\it Nuclear Physics Division,
Bhabha Atomic Research Centre  \\
Mumbai-400085, India }
\end {center}

\begin {abstract}
The cross section is estimated for the $\pi^+\pi^-\pi^0$ invariant mass
distribution in the $\gamma p$ reaction in the GeV region. This reaction
is assumed to proceed through the formation of the $\omega$ meson in
the intermediate state, since the production cross section for this meson
in the $\gamma p$ reaction in GeV region is significant and it has large
branching ratio $(88.8\%)$ in the $\pi^+\pi^-\pi^0$ channel. The cross
sections for this reaction have been calculated using the energy dependent
reaction amplitude, i.e., $f_{\gamma p \to \omega p}(0)$, extracted from
the latest $\omega$ meson photoproduction data. We use established
procedure to calculate other factors, like width and propagator of the
$\omega$ meson, so that our calculation can provide reliable cross
section. The calculated results reproduce the measured $\pi^+\pi^-\pi^0$
invariant mass distribution spectra in the $\gamma p$ reaction.
\end {abstract}
PACS number(s): 25.20.Lj

\section {Introduction}
\label {Int}

~~~~
The vector meson production in the nuclear and particle reactions shows
growing interest over the years, since it revealed many rewarding physics
in various topics of the hadronic physics. The importance of the vector
meson in context to the pion production in the GeV region was realized long
back \cite{vmp1, vmp2, vmp3}. The dilepton production in this energy region
is undoubtly explained due to the production of the vector mesons in the
intermediate state \cite{vsn1, vsn2}.
The vector dominance model (VDM) gives special status to the
vector meson in describing the electromagnetic interaction between the
lepton and hadron \cite{vdm}. The vector meson can probe the low-lying
nucleonic resonances \cite{huber} since it couples to these resonances
through the tail of its mass distribution. To be added, the vector meson
production process can be used to search the missing resonances \cite{msrs}.
For the later purpose, the $\omega p$ system is a preferable choice because
of the narrow width (i.e., 8.43 MeV) of the $\omega$ meson. Of course, this
system has a restriction to identify the nucleonic resonance of isospin
$I=\frac{1}{2}$.

The vector meson has a significant role in understanding the quark-gluon
picture of the hadron, since the Quantum Chromodynamics (QCD) subtles it as
a spin-triplet bound state of the specific valence quark $(q)$ and
anti-quark $(\bar{q})$ pair in the sea of $q\bar{q}$ pairs of all flavors
including gluons. Indeed, the static quark model (a simplified picture
elucidating the vector meson as a spin-triplet bound state of specific
$q\bar{q}$ pair only) \cite{qbks} is deceptively successful to account
some properties for the vector meson, such as spin-isopin for this meson,
decay of vector meson, the potential energy between the $q\bar{q}$ pair,
.... etc. In the energy region of hard scattering, the fluctuation of the
quark-gluon configuration inside a hadron can be studied through the vector
meson production process. This fluctuation makes better transparency (called
color transparency) for the vector meson propagation through the nucleus
\cite{clrt}.

Recent past, CBELSA/TAPS collaboration at ELSA did experiment for the
$\omega$ meson photoproduction on proton and nuclear targets \cite{elsa}.
In this measurement, the $\omega$ meson was probed by the $\pi^0\gamma$
invariant mass distribution spectrum. We have studied the mechanism
for this reaction and calculated the cross section for it \cite{das}. To
be mentioned, the $\omega$ meson has only $8.5\%$ branching ratio in
the $\pi^0\gamma$ channel whereas it dominantly $(88.8\%)$ decays into the
$\pi^+\pi^-\pi^0$ channel. Therefore,
we calculate the cross section for the $\pi^+\pi^-\pi^0$ invariant mass
distribution (since it gives better signal for the $\omega$ meson production)
in the $\gamma p$ reaction. We assume that this reaction proceeds as
$ \gamma p \to \omega p $; $ \omega \to \pi^+\pi^-\pi^0 $. To describe the
$\omega$ meson photoproduction in the intermediate state, we extract the
energy dependent $ \gamma p \to \omega p $ reaction amplitudes from the
latest measurement of the four momentum transfer distribution (elaborated
later). We follow the widely used procedure to evaluate all other factors
appearing in the cross section (e.g., the propagator and width of the
$\omega$ meson) to get the reliable cross section for the
$\pi^+\pi^-\pi^0$ invariant mass distribution in the above mentioned
reaction.

The formalism for this reaction is described in sec.~2. The calculated
results (along with the data) have been described in sec.~3. The conclusion
of this study is presented in sec.~4.

\section {Formalism}
\label {for}

~~~~
The differential cross section for the $\pi^+\pi^-\pi^0$ invariant mass
distribution in the reaction:
$ \gamma p \to \omega p$; $ \omega \to \pi^+\pi^-\pi^0 $ can be expressed as
\begin{equation}
\frac{d\sigma (m,E_\gamma)}{dm}
= \int d\Omega_\omega [KF] \Gamma_{\omega \to 3\pi}(m)
  |G_\omega (m)|^2 | F(\gamma p \to \omega p^\prime) |^2,
\label{dsc1}
\end{equation}
where $[KF]$ represents the kinematical factor for the above reaction. It
is given by
\begin{equation}
[KF] = \frac{1}{(2\pi)^3} \frac{ k^2_\omega m_p m^2 }
{ k_\gamma | k_\omega (E_\gamma+m_p) - {\bf k_\gamma . {\hat k}_\omega}
E_\omega |}.
\label{kf}
\end{equation}
All symbols carry their usual meanings.

$ \Gamma_{\omega \to 3\pi} (m) $ in Eq.~(\ref{dsc1}) denotes the width for
$ \omega (\mbox{at rest}) \to \pi^+\pi^-\pi^0 $, i.e.,
$ \Gamma_{\omega \to 3\pi} (m)  \equiv
\Gamma_{\omega \to \pi^+\pi^-\pi^0} $, governed by the Lagrangian density
$ {\cal L}_{\omega 3\pi} = \frac{f_{\omega3\pi}}{m^3_\pi}
\epsilon_{\mu\nu\lambda\sigma} \omega^\mu k^\nu_{\pi^+}
k^\lambda_{\pi^-} k^\sigma_{\pi^0} $ \cite{saku}. The expression for
$ \Gamma_{\omega \to \pi^+\pi^-\pi^0} $ is given in Eq.~(\ref{wdth1}). The
$\omega$ meson propagator $ G^{\mu\mu^\prime}_\omega (m) $ is given by
$ G^{\mu\mu^\prime}_\omega (m)
= (-g^{\mu\mu^\prime}+\frac{1}{m^2}k^\mu_\omega k^{\mu^\prime}_\omega)
G_\omega (m) $.
$g^{\mu\mu^\prime}$ couples the $\pi^+\pi^-\pi^0$ field (in the final
state) to the vector field. The second part of the propagator does not
contribute because of the antisymmetric coupling of $\omega$ meson
to three pion (see in ${\cal L}_{\omega3\pi}$, written above).
The factor $G_\omega(m)$,
which also appears in Eq.~(\ref{dsc1}), describes the scalar part of the
$\omega$ meson propagator in the free space. The expression for it is
\begin{equation}
G_\omega (m) = \frac{ 1 }{ m^2-m^2_\omega + im_\omega \Gamma_\omega (m) },
\label{omega}
\end{equation}
with $ m_\omega \simeq 782 $ MeV. $\Gamma_\omega(m)$ denotes the total free
space decay width for the $\omega$ meson. It is composed of hadronic,
semi-hadronic and leptonic decay channels \cite{pdg1}:
\begin{equation}
\Gamma_\omega (m) \approx
  \Gamma_{\omega \to \pi^+\pi^-\pi^0} (88.8\%)
+ \Gamma_{\omega \to \pi^0\gamma} (8.5\%)
+ \Gamma_{\omega \to \pi^+\pi^-} (2.21\%)
+ \Gamma_{\omega \to l^+l^-} (\sim 10^{-4}\%).
\label{wdths}
\end{equation}
Since the leptonic decay channels (i.e., $\Gamma_{\omega \to l^+l^-}$) for
the $\omega$ meson are insignificant in compare to other decay channels,
they are ignored in this calculation.

The form for $ \Gamma_{\omega \to \pi^+\pi^-\pi^0} (m) $ in
Eq.~(\ref{wdths}), as shown by Sakurai \cite{saku}, is given by
\begin{equation}
\Gamma_{\omega \to \pi^+\pi^-\pi^0} (m)
=\Gamma_{\omega \to \pi^+\pi^-\pi^0} (m_\omega)
 \frac{m}{m_\omega} \frac{(m-3m_\pi)^4}{(m_\omega-3m_\pi)^4}
 \frac{U(m)}{U(m_\omega)},
\label{wdth1}
\end{equation}
with $ \Gamma_{\omega \to \pi^+\pi^-\pi^0} (m_\omega \simeq 782 ~\mbox{MeV})
\simeq 7.49 $ MeV \cite{pdg1}. The function $U(m)$ is described in
Ref.~\cite{saku} as $ U(m) \to 1 $ for $ m \to 3m_\pi $ and
$ U(m) \to 1.6 $ for $ m \to 787 $ MeV. It is also taken equal to 1.6 for
$ m > 787 $ MeV.

The width $ \Gamma_{\omega \to \pi^0\gamma} (m) $ appearing in
Eq.~(\ref{wdths}) is evaluated using the Lagrangian density:
$ {\cal L}_{\omega\pi^0\gamma} = \frac{ f_{\omega\pi\gamma} }{ m_\pi }
\epsilon_{\mu\nu\rho\sigma} \partial^\mu A^\nu \pi^0 \partial^\rho
\omega^\sigma $ \cite{dazi}. The expression for it is
\begin{equation}
\Gamma_{\omega \to \pi^0\gamma} (m)
=\Gamma_{\omega \to \pi^0\gamma} (m_\omega)
 \left [ \frac{k(m)}{k(m_\omega)} \right ]^3,
\label{wdth2}
\end{equation}
with $ \Gamma_{\omega \to \pi^0\gamma} (m_\omega) \simeq 0.72$ MeV at
$m_\omega \simeq 782$ MeV \cite{pdg1}. $k(m)$ denotes the momentum of pion
originating due to the $\omega$ meson of mass $m$ decaying at rest.

In Eq.~(\ref{wdths})$, \Gamma_{\omega \to \pi^+\pi^-} (m) $ denotes the
width for the $\omega$ meson decaying to $\pi^+\pi^-$ channel.  In fact,
this channel arises due to small isovector component present in the
physical $\omega$ meson, i.e.,
$ \omega = \omega_I(0,0) + \epsilon \rho_I (1,0) $ \cite{vsn2}. Here,
$\omega_I(0,0)$ and $\rho_I (1,0)$ denote the isoscaler and isovector
fields respectively. $\epsilon$ is the small mixing parameter.
Therefore, the
isovector $\pi^+\pi^-$ current can strongly couple to $\rho_I (1,0)$
in the Lagrangian density ${\cal L}_{\omega\pi\pi}
= f_{\omega \pi\pi} ( \vec {\pi} \times \partial_\mu \vec {\pi}) \cdot
{\bf \omega}^\mu $, and $\omega_I(0,0)$ can be ignored for this purpose.
We do not use isovector sign on the $\omega$ meson appearing in
the Lagrangian since the $\rho_I (1,0)$ content in the $\omega$ meson is
very small.
To be mentioned, it has been shown clearly in the Ref.~\cite{vsn2} that
even if $ \omega_I(0,0) \to \pi^+\pi^- $ is allowed due to charge symmetry
violation (CSV), the $\pi^+\pi^-$ emission from the $\omega$ meson is
possible only due to $ \rho_I(1,0) \to \pi^+\pi^- $ for
$ m_\rho \sim m_\omega $ and $ \Gamma_\rho >> \Gamma_\omega $.
The width for the $ \omega \to \pi^+\pi^- $ channel is worked out as
\begin{equation}
\Gamma_{\omega \to \pi^+\pi^-} (m)
=\Gamma_{\omega \to \pi^+\pi^-} (m_\omega) \frac{m_\omega}{m}
 \left [ \frac{k(m)}{k(m_\omega)} \right ]^3.
\label{wdth3}
\end{equation}
The value for $ \Gamma_{\omega \to \pi^+\pi^-}
(m_\omega \simeq 782 ~\mbox{MeV}) $, according to Ref.~\cite{pdg1}, is
approximately equal to 0.19 MeV. $k(m)$ represents the pion momentum in the
$\pi^+\pi^-$ cm of system.

The $\omega$ meson dominantly decays to $\pi^+\pi^-\pi^0$ channel (see
$\Gamma_\omega (m)$ in Eq.~(\ref{wdths})). Therefore, the mass
distribution of the $\omega$ meson is significantly governed by the decay
width of this channel, i.e., $\Gamma_{\omega \to \pi^+\pi^-\pi^0} (m)$,
expressed in Eq.~(\ref{wdth1}). The $\omega$ meson possesses narrow width
(about 8.43 MeV) in the free space. To justify it, we plot the mass $m$
dependence of $\Gamma_\omega (m)$ in Eq.~(\ref{wdths}) as well as
$\Gamma_{\omega \to \pi^+\pi^-\pi^0} (m)$ in Eq.~(\ref{wdth1}) in
Fig.~1. This figure shows that expressions used for these widths
duly reproduce the respective measured values at $m=m_\omega$, quoted in
Ref.~\cite{pdg1}.

The generalised potential for the $\omega$ meson photoproduction in the
nuclear reaction can be expressed as
$ F (\gamma p \to \omega p^\prime) \varrho({\bf r}) $ \cite{gnpt},
where $ \varrho({\bf r}) $ represents the density distribution
of the target nucleus. For the point particle (i.e., proton target), the
density distribution for it is $ \varrho({\bf r}) = \delta ({\bf r}) $.
The form for $ F (\gamma p \to \omega p^\prime) $, which appears in
Eq.~(\ref{dsc1}), in the center of mass system is given by
\begin{equation}
F (\gamma p \to \omega p^\prime)
= -4\pi \left [ 1 + \frac{E_\omega}{E_{p^\prime}} \right ]
f_{\gamma p \to \omega p^\prime} (0),
\label{fgo1}
\end{equation}
where $ f_{\gamma p \to \omega p^\prime} (0) $ is the forward amplitude for
the $ \gamma p \to \omega p^\prime $ reaction. In the cross section
in Eq.~(\ref{dsc1}), $ f_{\gamma p \to \omega p^\prime} (0) $ appears
in the form of $| f_{\gamma p \to \omega p^\prime} (0) |^2$ which is
related to the four momentum $q^2$ transfer distribution
$ d\sigma (\gamma p \to \omega p^\prime) / dq^2 $ \cite{shmt} as
\begin{equation}
\frac{d\sigma}{dq^2} ( \gamma p \to \omega p^\prime ; q^2=0 )
=\frac{\pi}{k^2_\gamma} | f_{\gamma p \to \omega p^\prime} (0) |^2.
\label{fgo2}
\end{equation}
The forward $ d\sigma (\gamma p \to \omega p^\prime) / dq^2 $ is used to
obtain from the extrapolation of the measured
$ d\sigma (\gamma p \to \omega p^\prime) / dq^2 $.
In fact, the energy dependent values for it are reported in
Refs.~\cite{shmt, stk} for $ E_\gamma \ge 1.6 $ GeV. In the present study,
we deal with the $\omega$ meson photoproduction for the beam energy range:
$ E_\gamma (\mbox{GeV}) = 1.1 - 9.3 $. In the lower energy region, i.e.,
$ E_\gamma \le 2.6 $ GeV, the data for the
$ d\sigma (\gamma p \to \omega p^\prime) / dq^2 $ distribution are taken
from the measurement (done in recent past) with the SAPHIR detector at
electron stretcher ring (ELSA), Bonn \cite{bar1}.
In this measurement, they have reported the measured
$ d\sigma (\gamma p \to \omega p^\prime) / dq^2 $ vs $ | q^2 - q^2_{min} | $
($q^2_{min}$ is defined in Ref.~\cite{shmt, tmn}). Therefore, we extract
$ | f_{\gamma p \to \omega p^\prime} (0)|^2 $ from the SAPHIR data for
$ E_\gamma \le 2.6 $ GeV. For $ E_\gamma \ge 2.6 $ GeV, the energy dependent
$ | f_{\gamma p \to \omega p^\prime} (0)|^2 $ is evaluated from the forward
$ d\sigma (\gamma p \to \omega p^\prime) / dq^2 $ given in
Refs.~\cite{shmt, stk}.

The Eq.~(\ref{dsc1}) can be used to calculate the differential cross section
for the $\pi^+\pi^-\pi^0$ invariant mass distribution
$ \frac{d\sigma (m,E_\gamma)}{dm} $ due to $ \omega \to \pi^+\pi^-\pi^0 $
for a fixed beam energy $E_\gamma$. To describe this reaction for the
$\gamma$ beam of certain energy range, as it happens for the tagged photon,
we modulate the cross section given in Eq.~(\ref{dsc1}) with the beam
profile function $W(E_\gamma)$ \cite{kho}, i.e.,
\begin{equation}
\frac{d\sigma (m)}{dm} = \int^{E_\gamma^{mx}}_{E_\gamma^{mn}}
dE_\gamma W(E_\gamma) \frac{d\sigma (m,E_\gamma)} {dm}.
\label{dsc2}
\end{equation}
The profile function $W(E_\gamma)$ for the $\gamma$ beam, originating due
to the bremsstrahlung radiation of the electron, varies as
$W(E_\gamma) \propto \frac{1}{E_\gamma}$ \cite{kho}.

\section {Results and Discussion}
\label {rsld}
~~~~
We calculate the cross sections $\frac{d\sigma}{dm}$ for the
$\omega (\to \pi^+\pi^-\pi^0)$ meson mass distribution spectra using the
Eq.~(\ref{dsc1}) for beam energies ($E_\gamma$ in GeV) taken equal to 2.8,
4.7 and 9.3. The calculated results (solid curves) are compared in
Fig.~2 with the $\pi^+\pi^-\pi^0$ invariant mass distribution
spectra (presented by histograms) measured by Ballam et al., \cite{vmp2}.
In this figure, the calculated cross section is normalized to the measured
spectrum at the peak.
The sharp peak appearing in the calculated spectrum at $\sim 782$ MeV is the
characteristic feature for the $\omega$ meson. The calculated spectra, as
shown in this figure, reproduce well the respective peak positions of all
measured distributions. These agreements elucidate the production of the
$\omega$ meson in the intermediate state. The mismatch in widths between
the calculated and measured spectra (even in other figures also) can be
presumed due to resolution width associated with the detector. The
calculated cross section at the peak is increased to $\sim 10.51$ mb/GeV
at $E_\gamma=9.3$ GeV from 7.25 mb/GeV at $E_\gamma(\mbox{GeV})=2.8$ GeV.

We present in Fig.~3 the calculated cross section, i.e.,
$\frac{d\sigma}{dm}$ due to Eq.~(\ref{dsc2}), along with the data for
$\pi^+\pi^-\pi^0$ invariant mass distribution in the energy bins:
$E_\gamma$ (in GeV) = $ 1.1-1.5;
1.5-1.8; 1.8-2.5 ~ \mbox{and} ~ 2.5-6.0 $. In this figure, the histograms
represent the $\pi^+\pi^-\pi^0$ invariant mass distribution spectra measured
by the BUBBLE chamber group \cite{vmp3}. The dashed curves describe the
phase-space distributions. The solid curve in each energy bin
is obtained multiplying the calculated cross section by a normalizing factor
and adding it with the respective phase space. The normalizing factors
are found equal to 14.5, 2.9, 3.4 and 3.2 for the beam energy $(E_\gamma)$
bin:
$1.1-1.5$ GeV, $1.5-1.8$ GeV, $1.8-2.5$ GeV and $2.5-6$ GeV respectively.
This figure shows that the calculated result reproduces the peak position
in each energy bin. The actual magnitudes of the calculated cross sections
are presented in Fig.~4. This figure shows that the peak cross
section is enhanced to $\sim 7.73$ mb/GeV at
$ E_\gamma (\mbox{GeV}) = 2.5-6.0 $ from $\sim 2.31$ mb/GeV at
$ E_\gamma (\mbox{GeV}) = 1.1-1.5 $.

Recent past, the SAPHIR collaboration measured the $\pi^+\pi^-\pi^0$
invariant mass distribution spectrum in the energy bin:
$ 1.2 < E_\gamma (\mbox{GeV}) < 1.25 $ with an additional constraint:
$ 0.3 < q^2_{min}-q^2 (\mbox{GeV}^2) < 0.4 $ \cite{bar1}. Since the
variation in the beam energy $E_\gamma$ is negligibly small (less than
50 MeV) in compare to the energy in the bin, we calculate
$\frac{d\sigma}{dm}$ using Eq.~(\ref{dsc1}) at $E_\gamma=1.225$ GeV for
this spectrum.
We compare the calculated result with the measured spectrum (stated
above) in Fig.~5. In this figure, the measured distribution is
shown by the histogram whereas the solid curve represent the calculated
spectrum (normalised to the peak of the measured distribution). The
magnitude of the calculated cross section at the peak is about 1.7 (mb/GeV).
In this case also, the calculated result is well accord with the measured
distribution.

\section {Conclusions}
\label {cncls}
~~~~
We have calculated the differential cross section for the $\pi^+\pi^-\pi^0$
invariant mass distribution in the $\gamma p$ reaction in the GeV region.
Since the $\omega$ meson couples strongly to $\pi^+\pi^-\pi^0$ in this
energy region, we consider that this event in the final state arises due
to the decay of the $\omega$ meson produced in the intermediate state. The
agreement between the calculated and measured peak positions corroborates
this consideration. The reaction amplitude $ f_{\gamma p \to \omega p} $,
which is extracted from the latest $\omega$ meson photoproduction data,
is used to estimate the magnitude of the cross section. Other factors
in this calculation are evaluated using the well known procedure for
them. Therefore, our calculation gives reliable cross section for the
$\pi^+\pi^-\pi^0$ invariant mass distribution in the $\gamma p$ reaction.

\section {Acknowledgements}
\label {ackn}
~~~~
I gratefully acknowledge A. K. Mohanty, R. K. Choudhury and S. Kailas.

\newpage
{\bf Figure Captions}
\begin{enumerate}

\item
The dependence of $\Gamma_\omega (m)$ in Eq.~(\ref{wdths})  and
$\Gamma_{\omega \to \pi^+\pi^-\pi^0} (m)$ in Eq.~(\ref{wdth1}) on the
$\omega$ meson mass.

\item
The calculated $ \omega (\to \pi^+\pi^-\pi^0) $ meson mass distribution
spectra for various beam (gamma) energies are presented. $m$ denotes the
mass of the $\omega$ meson, i.e., the $\pi^+\pi^-\pi^0 $ invariant mass
in the measurement.
The histograms (a) represent the $\pi^+\pi^-\pi^0$ invariant mass
distribution spectra (along with the background $< 10 \%$) measured by
Ballam et. al., \cite{vmp2}. They are given in Events/0.02 GeV.
The solid curves (b) show the calculated results, i.e.,
$\frac{d\sigma}{dm}$ due to the Eq.~(\ref{dsc1}). The calculated results
are normalized to the measured spectra at the respective peaks.

\item
The calculated $ \omega (\to \pi^+\pi^-\pi^0) $ meson mass distribution
spectra are compared with the data due to BUBBLE chamber group \cite{vmp3}.
In each energy bin, the histogram (a) represents the number of counts for the
measured $\pi^+\pi^-\pi^0$ invariant mass distribution spectrum and the
dashed curve describes the phase space. The solid curves (b) are related to
the calculated cross section $\frac{d\sigma}{dm}$ in Eq.~(\ref{dsc2}),
explained in the text.

\item
The calculated $ \omega (\to \pi^+\pi^-\pi^0) $ meson mass distribution
spectra for various beam (gamma) energy bins have been presented. The curves
appearing in this figure are the calculated results due to Eq.~(\ref{dsc2}).
This figure distinctly shows the enhancement in the cross section with the
beam energy.

\item
The calculated $ \omega (\to \pi^+\pi^-\pi^0) $ meson mass distribution
spectrum is compared with the data due to SAPHIR collaboration \cite{bar1}.
The histogram (a) represents the measured number of events for the
$\pi^+\pi^-\pi^0$ invariant mass distribution spectrum \cite{bar1}. The solid
curve (b) corresponds to the calculated cross section ($\frac{d\sigma}{dm}$ in
Eq.~(\ref{dsc1})), normalized to the peak of the measured distribution.

\end{enumerate}

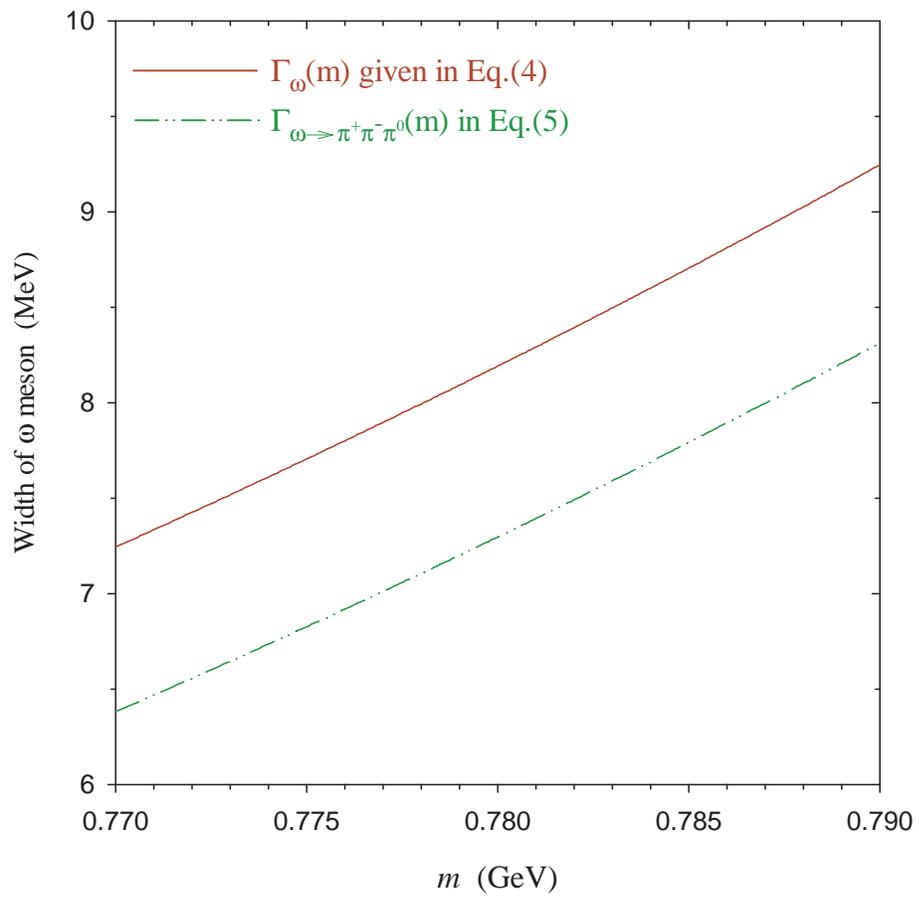
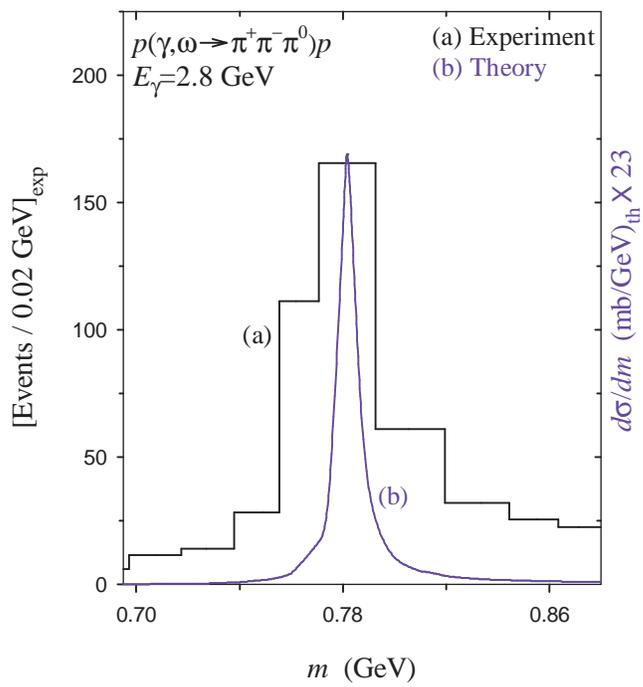
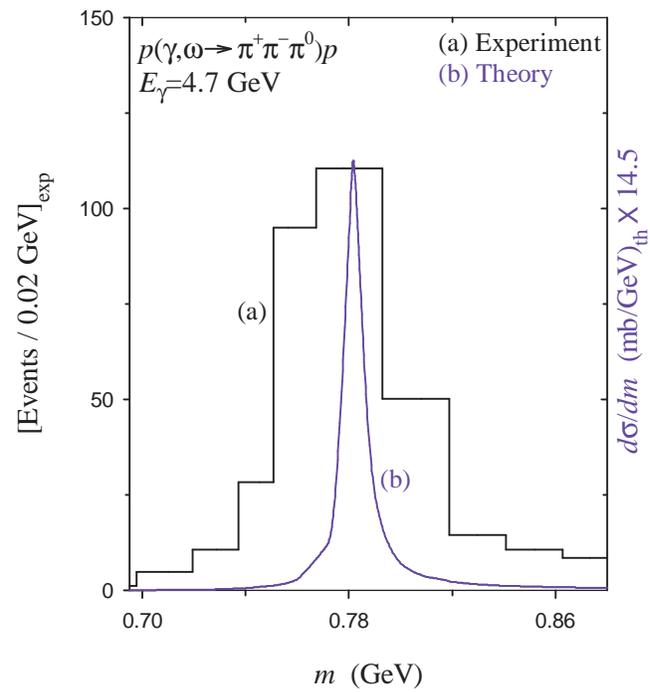
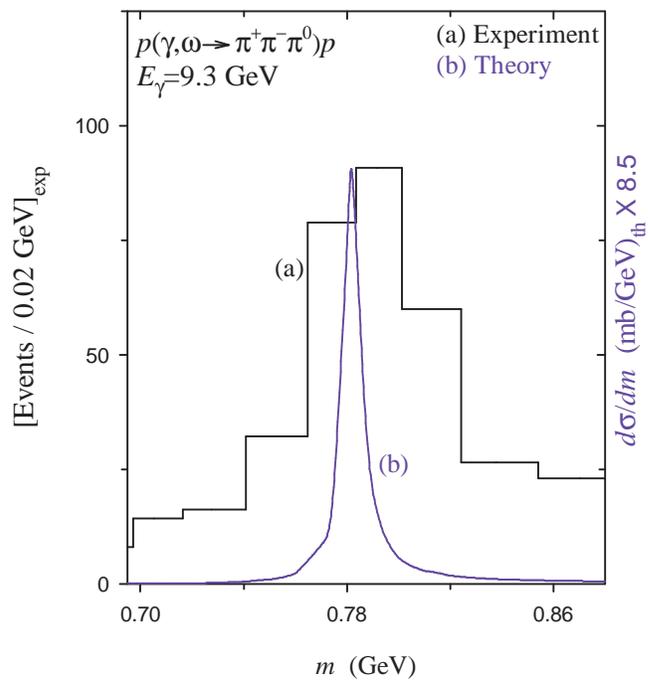
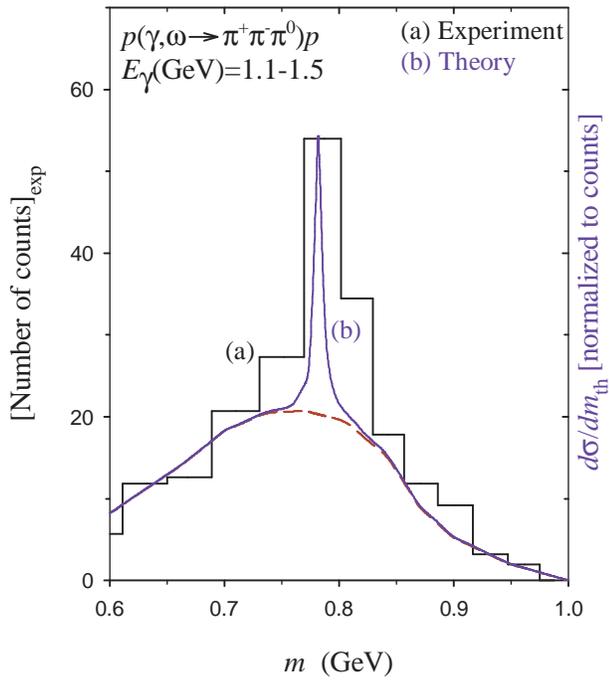
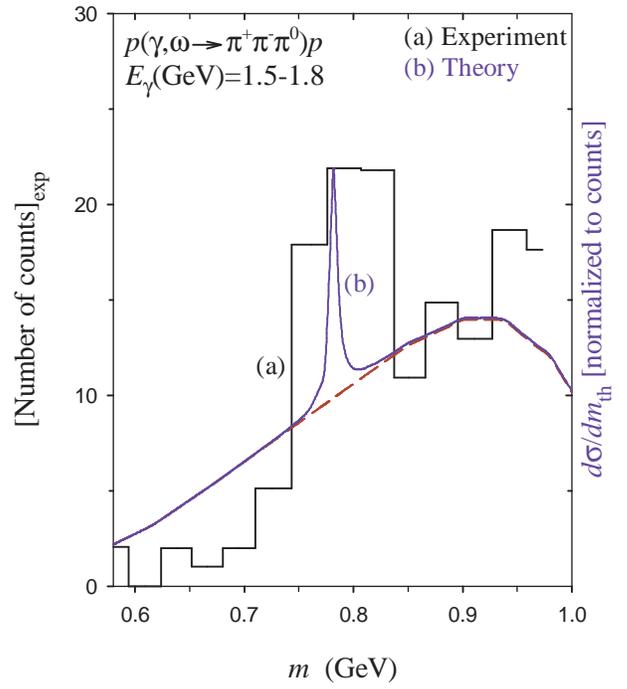
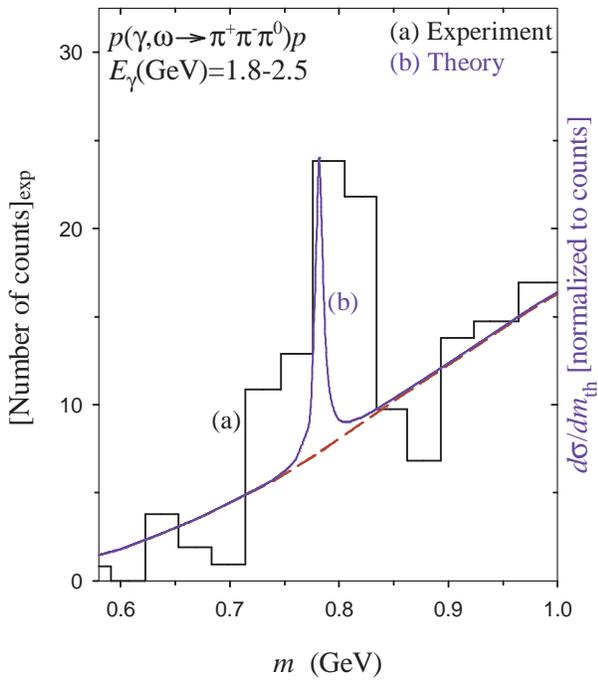
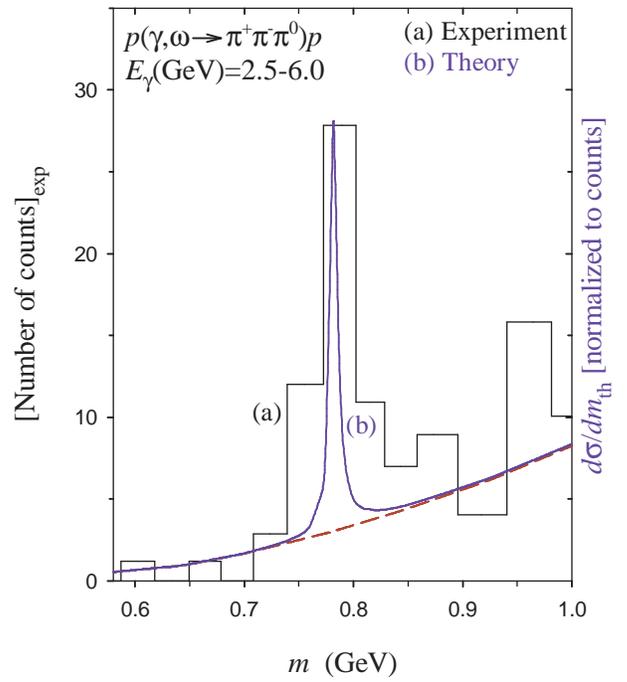
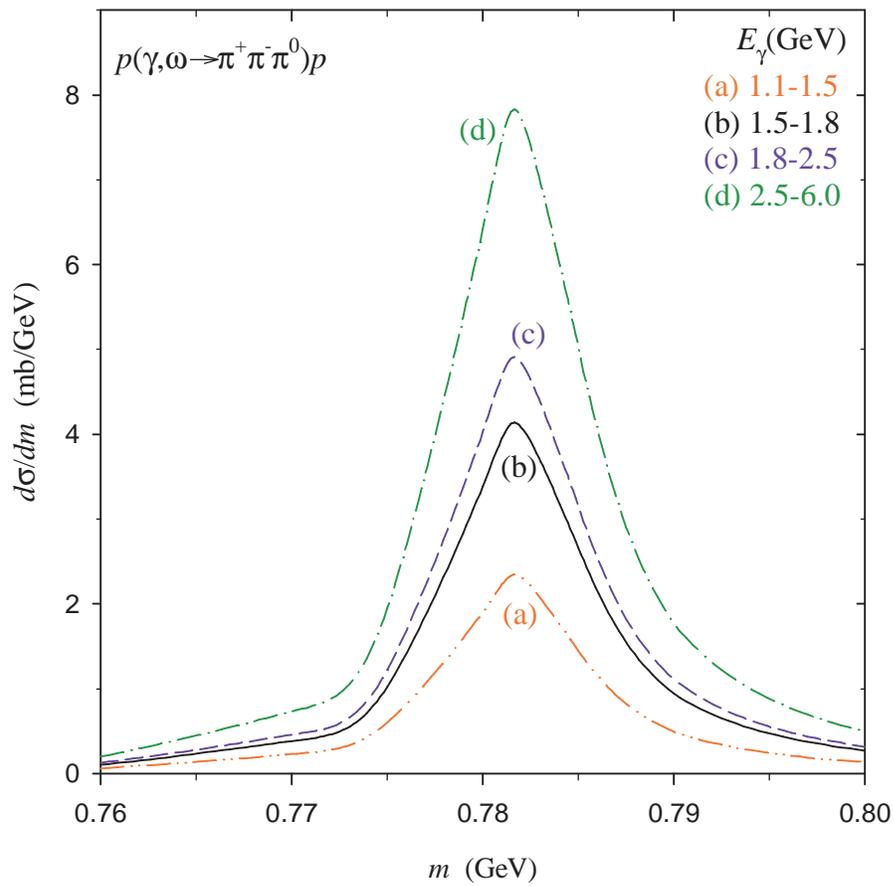
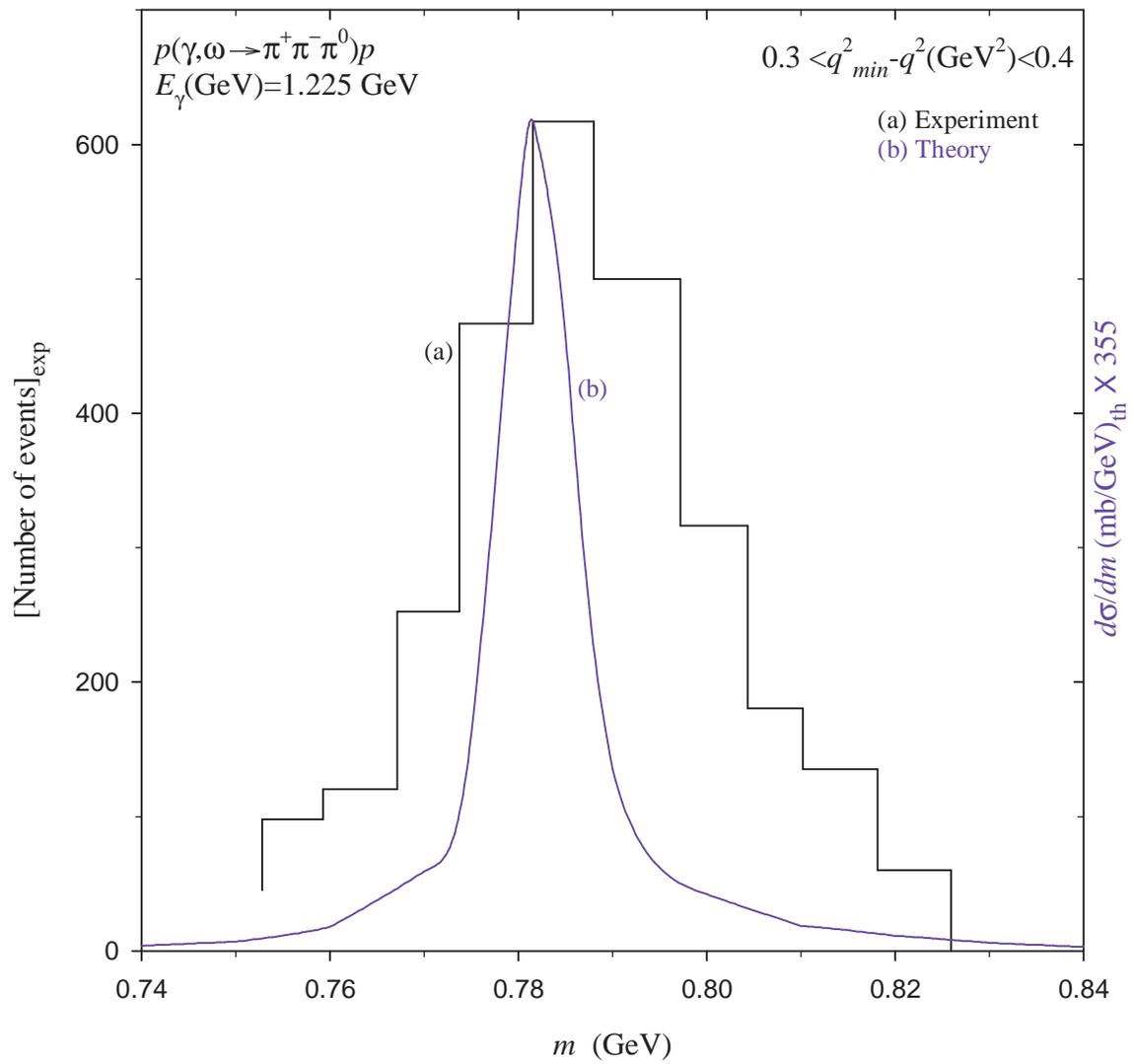
\end{document}